\documentclass[preprintnumbers,amsmath,amssymb]{revtex4}

\usepackage{graphicx}
\usepackage{dcolumn}
\usepackage{bm}

\begin{document}
\preprint{}

\title{Controlled quantum teleportation and secure direct communication}

\author{ T Gao$^{1,2,3}$, F L Yan $^{1,4}$ and Z X Wang $^3$}

\affiliation { $^1$ ~CCAST (World Laboratory), P.O. Box 8730, Beijing 100080, China\\
 $^2$ College of Mathematics and Information Science, Hebei Normal University, Shijiazhuang 050016, China
\\
$^3$  Department of Mathematics, Capital Normal University, Beijing 100037, China\\
$^4$ College of Physics, Hebei Normal University, Shijiazhuang 050016, China}

\begin{abstract}
 We present a controlled quantum teleportation protocol. In the protocol, quantum information of an
unknown state of a 2-level particle is faithfully transmitted from a sender (Alice) to a remote receiver (Bob)
via an initially shared  triplet  of entangled particles under the control of the supervisor Charlie. The
distributed entangled particles shared by Alice, Bob and Charlie function as a quantum information channel for
faithful transmission. We also propose a controlled and secure direct communication scheme by means of this
teleportation. After insuring the security of the quantum channel, Alice encodes the secret message directly on
a sequence of  particle states and transmits them to Bob supervised by Charlie using this controlled quantum
teleportation. Bob can read out the encoded message directly by the measurement on his qubit. In this scheme,
the controlled quantum teleportation transmits Alice's message without revealing any information to a potential
eavesdropper. Because there is not a transmission of the qubit
 carrying the secret message between Alice and Bob in the public channel, it is
completely secure for controlled and  direct secret communication if perfect quantum channel is used. The
feature of this scheme is that the communication between two sides depends on the agreement of the third side.
\end{abstract}


 \maketitle

\leftskip 0cm {\noindent \bf 1. Introduction}\\

{\noindent It is well-known that cryptography plays an increasing  important role in the whole  world. The goal
of secure communication is to make two distant parties to  exchange information  but not to allow the third
party to steal or alter the content of information. Before transmitting  their secret messages the two distant
parties must distribute secret key first. Since it is difficult to distribute secret key through a classical
channel, people have paid  a lot of attention to quantum key distribution. In 1984, Bennett and Brassard
presented the first quantum cryptography  protocol [1]. Up to now, there have already been several  quantum key
distribution schemes such as in Refs.[2-16].}

Recently, a novel quantum direct communication protocol has been proposed [17] that allows secure direct
communication, i.e. the two parties   communicate important messages directly without first establishing  a
shared secret key  to encrypt them and the message is deterministically sent through the quantum channel, but
can only be decoded after a final transmission of classical information. Bostr\"{o}m and Felbinger [18] put
forward a direct communication scheme, the "ping-pong protocol", which is insecure if it is operated in a noisy
quantum channel, as indicated  by W\'{o}jcik  [19]. More recently, Deng et al. [20] suggested a two-step quantum
direct communication protocol using Einstein-Podolsky-Rosen pair block. However in all these secure direct
communication schemes it is necessary to send the
  qubits with  secret messages in the public channel. Therefore,
  Eve can  attack the qubits  in
   transmission. Yan and Zhang [21] presented  a scheme for secure direct
   and confidential communication between Alice and Bob, using
   Einstein-Podolsky-Rosen pairs and teleportation
    [22]. We proposed a protocol for controlled and secure  direct communication using GHZ state and teleportation
    [23].   Because  there is not a transmission of  the qubits
    carrying the secret messages between Alice
and Bob in the public channel, they are completely secure for direct secret communication if perfect quantum
channel is used.

  Quantum teleportation  is  commonly considered as one of the most striking
progress of quantum information theory. In the original quantum teleportation scheme of Bennett et al. [22],
quantum information of an unknown quantum state of a d-level particle is faithfully transmitted from a sender
(Alice) to a spatially distant receiver (Bob) via an initially shared  maximally entangled state with the help
of classical communication. To date teleportation has been generalized to many cases [24-37]. In this paper, we
present a controlled quantum teleportation protocol, in which the third  side (the supervisor Charlie) is
included, so that the quantum channel is controlled by this additional side, and  quantum information of an
unknown quantum state of a 2-level particle cannot be transmitted unless all three sides agree to cooperate. We
also propose  another scheme for controlled secure direct communication with the aid of a   three-particle
entangled state and this controlled teleportation.  After insuring the security of the quantum channel, Alice
encodes the secret messages directly on a sequence of particle  states and transmits them to Bob supervised by
Charlie using this controlled quantum teleportation. Bob can read out the encoded messages directly by the
measurement on his qubit. In this scheme, the controlled quantum teleportation transmits Alice's message without
revealing any information to a potential eavesdropper. Because there is not a transmission of the qubit
 carrying the secret messages between Alice and Bob in the public channel, it is
completely secure for controlled and  direct secret communication if perfect quantum channel is used. The
feature of this scheme is that the communication between two sides depends on the agreement of the third side.\\

The rest of the paper is organized as follows. In section 2 we present a protocol for controlled quantum
teleportation. In the protocol, quantum information of an unknown state of a 2-level particle is faithfully
transmitted from a sender (Alice) to a remote   receiver (Bob) via an initially shared  triplet in an entangled
state under the control of the supervisor Charlie. The distributed entangled particles shared by Alice, Bob and
Charlie function as a quantum information channel for faithful transmission. In section 3 we also propose a
controlled and secure direct communication scheme by means of this teleportation. A summary is given in Section
4.\\

{\noindent\bf 2. Controlled quantum teleportation}\\

{\noindent In Ref. [37], Zhou et al. proposed a controlled quantum teleportation scheme, where the entanglement
property of GHZ state is utilized. Here we present another controlled quantum teleportation protocol in which  a
triplet in an entangled state  (different from GHZ state) initially shared by the sender Alice,  the receiver
Bob and the supervisor Charlie functions as a quantum channel with the help of a classical information channel.}

 We suppose that the sender Alice,  the receiver Bob and the supervisor
Charlie are spatially separated from each other. The original unknown quantum state of a 2-level particle  Alice
wishes to teleport  to Bob is
\begin{equation}\label{m}
   |\psi\rangle_M=a|0\rangle_M+b|1\rangle_M,
\end{equation}
where $|a|^2+|b|^2=1$. The quantum channel initially shared by Alice, Bob and Charlie  is a three-particle
entangled state:
\begin{equation}\label{abc}
  |\xi\rangle_{ABC}=\frac{1}{2}(|000\rangle+|110\rangle+|011\rangle+|101\rangle)_{ABC}.
\end{equation}
The joint state of the system is
\begin{equation}\label{s}
|\psi\rangle_M|\xi\rangle_{ABC}=(a|0\rangle+b|1\rangle)_M\otimes\frac{1}{2}(|000\rangle+|110\rangle
+|011\rangle+|101\rangle)_{ABC}.
\end{equation}

If  Charlie would like to help Alice for the quantum teleportation, he should just measure his portion of $ABC$,
namely qubit $C$, on the basis $ \{|0\rangle_C, |1\rangle_C\}$, and transfer the result of his measurement to
Alice and Bob via a classical channel. Then the particles $M$, $A$ and $B$ will become one of the states:
\begin{eqnarray}
|\phi_1\rangle_{MAB} = _C\langle 0|\psi\rangle_M|\xi\rangle_{ABC}=
(a|0\rangle+b|1\rangle)_M\otimes\frac{1}{\sqrt{2}}(|00\rangle+|11\rangle)_{AB},\\
|\phi_2\rangle_{MAB} = _C\langle 1|\psi\rangle_M|\xi\rangle_{ABC}=
(a|0\rangle+b|1\rangle)_M\otimes\frac{1}{\sqrt{2}}(|01\rangle+|10\rangle)_{AB},
\end{eqnarray}
which can be rewritten as
\begin{eqnarray}
  |\phi_1\rangle_{MAB}=\frac{1}{2}[|\Phi^+\rangle_{MA}(a|0\rangle+b|1\rangle)_B
                         +|\Psi^+\rangle_{MA}(a|1\rangle+b|0\rangle)_B
                         +|\Phi^-\rangle_{MA}(a|0\rangle-b|1\rangle)_B
                         +|\Psi^-\rangle_{MA}(a|1\rangle-b|0\rangle)_B],\\
  |\phi_2\rangle_{MAB}=\frac{1}{2}[|\Phi^+\rangle_{MA}(a|1\rangle+b|0\rangle)_B
                           +|\Psi^+\rangle_{MA}(a|0\rangle+b|1\rangle)_B
                           +|\Phi^-\rangle_{MA}(a|1\rangle-b|0\rangle)_B
                           +|\Psi^-\rangle_{MA}(a|0\rangle-b|1\rangle)_B].
\end{eqnarray}
Here $ |\Phi^\pm\rangle_{MA}=\frac {1}{\sqrt 2}(|00\rangle\pm|11\rangle)_{MA}, |\Psi^\pm\rangle_{MA}=\frac
{1}{\sqrt 2}(|01\rangle\pm|10\rangle)_{MA}$. Now Alice performs a Bell state measurement on her two particles
$M$ and $A$ and sends the result to Bob through a classical communication channel. Depending on both  Charlie
and Alice's  eight possible measurement outcomes  $|0\rangle_C|\Phi^+\rangle_{MA}$,
 $|0\rangle_C|\Psi^+\rangle_{MA}$, $|0\rangle_C|\Phi^-\rangle_{MA}$, $|0\rangle_C|\Psi^-\rangle_{MA}$,
$|1\rangle_C|\Phi^+\rangle_{MA}$, $|1\rangle_C|\Psi^+\rangle_{MA}$, $|1\rangle_C|\Phi^-\rangle_{MA}$ and
$|1\rangle_C|\Psi^-\rangle_{MA}$, Bob applies the corresponding  unitary operators $I$, $X$, $Z$,  $ZX$, $X$,
$I$, $ZX$ and $Z$, respectively, on his particle in order to recover the initial state
$|\psi\rangle_B=a|0\rangle_B+b|1\rangle_B$. Here $I=|0\rangle\langle 0|+|1\rangle\langle1|$, $X=|0\rangle\langle
1|+|1\rangle\langle 0|$, $Z=|0\rangle\langle 0|-|1\rangle\langle1|$. So an initial state $|\psi\rangle_M$ is
faithfully transmitted to Bob's particle via the quantum channel.

The feature of this scheme is that teleportation between two sides depends on the agreement of the third side.
It
is therefore another kind  "Controlled Quantum Teleportation" different from that in [37].\\

{\noindent\bf 3. Controlled and secure direct communication}\\

{\noindent Next we present a controlled and secure direct communication scheme (different from that in [23]) by
the above controlled  teleportation. The feature of this scheme is that the communication between two sides
depends on the agreement of the third side.}

Consider the scenario. Suppose that the administrative personnel of server,  Charlie, wishes to control the
correspondence between users. This means that if and only if attaining his permissibility, one  can correspond
with another. However, the users want their communication to be secret to Charlie and not altered during
transmission. What they can do? Our following scheme will suit this task. In our scheme, there are three
parties: the controller Charlie and the users Alice and Bob. How does this new scheme work?

1. Preparing quantum channel

Suppose that Alice,  Bob and Charlie share  a set of triplets of qubits  in an
 entangled  state $|\xi\rangle_{ABC}$ which function as quantum channel.
 Obtaining these triplets in an entangled   state could have come about in many different ways;
  for example, Charlie   prepares a sequence of  three-particle in an entangled state $|\xi\rangle_{ABC}$
  and then sends particles $A$ to Alice and particles $B$ to Bob, and keeps the corresponding particles $C$.
    Alternatively,
 a fourth  party could prepare  an ensemble of particles in $|\xi\rangle_{ABC}$, and ask Alice, Bob and Charlie
  to  each
 take a particle ($A$, $B$, $C$, respectively) in each triplet.  Or they could have met
 a long time ago and shared them,
 storing them until the present.  Alice, Bob and Charlie  then choose  randomly a subset of qubits
 in an entangled   state $|\xi\rangle_{ABC}$,
 and do   some  appropriate tests of fidelity.  Passing the test certifies that
  they continue to hold sufficiently pure, entangled quantum states.  However,
 if tampering has occurred,
 Alice, Bob and Charlie discard  these triplets, and  a new sequence of qubits in $|\xi\rangle_{ABC}$
  should be reproduced.

2. Secure direct communication using the above controlled teleportation

After insuring the security of the quantum channel, they begin  controlled and secure direct communication.

First Alice makes his particle sequence in the states, composed of $|+\rangle$ and $|-\rangle$, according to the
message sequence. For example if the message to be transmitted is 101001, then the sequence of particle states
should be in the state $|+\rangle|-\rangle|+\rangle|-\rangle|-\rangle|+\rangle$, i.e. $|+\rangle$ and
$|-\rangle$ correspond to 1 and 0 respectively. Here
\begin{equation}\label{i}
|+\rangle=\frac {1}{\sqrt 2}(|0\rangle+|1\rangle), ~~~~~~~~~~|-\rangle=\frac {1}{\sqrt 2}(|0\rangle-|1\rangle).
\end{equation}
The signal state carrying secret message that Alice wants to teleport is represented as
\begin{equation}
|\varphi\rangle_M=\frac {1}{\sqrt 2}(|0\rangle+b|1\rangle)_M,
\end{equation}
where $b=1$ and $b=-1$ correspond to $|+\rangle$ and $|-\rangle$ respectively.   The quantum state of the whole
system (the four qubits) reads
\begin{equation}
|\varphi\rangle_M|\xi\rangle_{ABC}=\frac {1}{\sqrt
2}(|0\rangle+b|1\rangle)_M\otimes\frac{1}{2}(|000\rangle+|110\rangle +|011\rangle+|101\rangle)_{ABC}.
\end{equation}
If Charlie allows the communications between the two users, he performs measurements on his qubit $C$ and
classically (and publicly) broadcasts the measurement outcome (on which basis of $\{|0\rangle_C, |1\rangle_C\}$
he obtained by projection) to the users Alice and Bob. After that, Alice makes a Bell state measurement on her
particles $M$ and $A$ and informs Bob of the  result  of measurement also publicly over a classical channel.
Once Bob has learned  the broadcast results of Charlie and Alice, Bob can 'fix up' his state, recovering the
signal state $|\varphi\rangle_B=\frac {1}{\sqrt 2}(|0\rangle+b|1\rangle)_B$, by applying appropriate quantum
gate. Then Bob measures  the basis $\{|+\rangle, |-\rangle\}$ and reads out the  messages that Alice wants to
transmit to him.

It is undeniable that this process of controlled quantum teleportation has similar notable features of the
original
 quantum  teleportation [22] which was mentioned in [21, 23]. For instance, the process
is entirely unaffected by any noise in the spatial environment between each other, and  the controlled
teleportation achieves perfect transmission of delicate information across a noisy environment and without even
knowing the locations of each other.  In the process Bob is left with a perfect instance of $|\varphi\rangle$
and hence no participants can gain any further information about its identity. So in our scheme controlled
quantum teleportation transmits Alice's message without revealing any information to a potential eavesdropper,
 if the quantum channel is  perfect  state in  Eq.(\ref{abc}) (perfect quantum channel).

The security  of this protocol only depends on the perfect quantum channel (pure  state in Eq.(\ref{abc})). Thus
as
long as the quantum channel is perfect, our scheme is absolutely reliable, deterministic and secure.\\

{\noindent\bf 4. Summary}\\

{\noindent In summary, we present a controlled quantum teleportation protocol. In the protocol, quantum
information of an unknown state of a 2-level particle is faithfully transmitted from a sender (Alice) to a
remote receiver (Bob) via an initially shared  triplet  in an entangled state under the control of the
supervisor Charlie. The distributed entangled particles shared by Alice, Bob and Charlie function as a quantum
information channel for faithful transmission. We also propose a controlled and secure direct communication
scheme by means of this controlled teleportation. After insuring the security of the quantum channel, Alice
encodes the secret messages directly on a sequence of particle states  and transmits them  to Bob by
teleportation supervised by Charlie. Evidently controlled teleportation transmits Alice's messages without
revealing any information to a potential eavesdropper. Bob can read out the encoded messages directly by the
measurement on his qubits. Because there is not a transmission of the qubit which carries the secret message
between Alice and Bob, it is completely secure for controlled and direct secret communication if perfect quantum
channel is used. The feature of this scheme is that the communication between two sides depends on the agreement
of the third side.}

Teleportation has been realized in the experiments [38, 39, 40], therefore our protocols for controlled quantum
teleportation and
secure direct communication will be realized by  experiment easily.\\

{\noindent\bf Acknowledgments}\\

{\noindent This work was supported by National Natural Science Foundation of China under Grant No.
10271081 and Hebei Natural Science Foundation under Grant No. A2004000141.}\\

{\parindent=0cm \bf References } \footnotesize
\begin{tabbing}
xxxxx\=1 \kill $[1]$   Bennett C H and Brassard G 1984  Proc. IEEE Int. Conf. on Computers, Systems and Signal
         Processing,  IEEE, New York, 175\\
$[2]$  Ekert A K 1991 {\it Phys. Rev. Lett.} {\bf  67}  661\\
$[3]$  Bennett C H, Brassard G and Mermin N D 1992  {\it Phys. Rev. Lett.} {\bf  68}  557\\
$[4]$  Bennett C H 1992  {\it Phys. Rev. Lett.} {\bf  68}  3121\\
$[5]$  Bennett C H and Wiesner S J 1992  {\it Phys. Rev. Lett.} {\bf   69}  2881\\
$[6]$  Goldenberg L and Vaidman L 1995  {\it Phys. Rev. Lett.} {\bf 75}  1239\\
$[7]$ Huttner B, Imoto N, Gisin N and Mor T 1995  {\it Phys. Rev.}  A {\bf 51}  1863\\
$[8]$   Koashi M and Imoto N   1997 {\it Phys. Rev. Lett.} {\bf 79}  2383\\
$[9]$   Bru$\ss$ D 1998 {\it Phys. Rev. Lett.} {\bf 81}  3018\\
$[10]$    Hwang W Y, Koh I G  and Han Y D 1998  {\it Phys.  Lett.} A {\bf 244}  489\\
$[11]$   Cabello A 2000  {\it Phys. Rev. Lett.} {\bf 85}  5635\\
$[12]$    Cabello A 2000 {\it Phys. Rev.} A {\bf 61}  052312\\
$[13]$  Long G L and Liu X S 2002 {\it Phys. Rev.} A {\bf 65}  032302\\
$[14]$   Xue P, Li C F and Guo G C 2002 {\it Phys. Rev.} A {\bf 65}  022317\\
$[15]$   Phoenix S J D, Barnett S M, Townsend P D and  Blow K J 1995 {\it J. Modern Optics} {\bf 42}  1155\\
$[16]$  Lo H K, Chan H F and Ardehali M  arXiv: quant-ph/0011056\\
$[17]$  Beige A et al 2002 {\it Acta Phys. Pol.} A {\bf 101}  357\\
$[18]$  Bostr\"{o}m K and Felbinger T 2002 {\it Phys. Rev. Lett.} {\bf 89}  187902\\
$[19]$  W$\acute{\rm o}$jcik A 2003 {\it Phys. Rev. Lett.} {\bf 90}  157901\\
$[20]$  Deng F G, Long G L and Liu X S 2003 {\it Phys. Rev.} A  {\bf 68}  042317\\
$[21]$  Yan F L and Zhang X Q  arXiv: quant-ph/0311132\\
$[22]$  Bennett C H et al 1993  {\it Phys. Rev. Lett.} {\bf 70}  1895 \\
$[23]$  Gao T arXiv: quant-ph/0312004\\
$[24]$  Murao M,  Plenio M B and  Vedral V 2000 {\it Phys. Rev.}  A {\bf 61}  032311\\
$[25]$   Ikram M,  Zhu S Y and  Zubairy M S  2000 {\it Phys. Rev.}  A {\bf 62}  022307 \\
$[26]$   Li W L,  Li C F and  Guo G C 2000 {\it Phys. Rev.} A {\bf 61}  034301\\
$[27]$  Gorbachev V N and   Trubilko  A I 2000 {\it J. Exp. Theor. Phys.} {\bf 91}  894\\
$[28]$   Lu H and  Guo  G C  2000 {\it Phys. Lett.}  A {\bf  276}  209\\
$[29]$   Zeng B,  Liu X S,  Li Y S and  Long G L 2002 {\it Commun. Theor. Phys.} {\bf 38}  537\\
$[30]$   Shi B S et al 2000 {\it Phys. Lett.} A {\bf  268}  161\\
$[31]$   Yan F L, Tan H G and  Yang L G 2002 {\it Commun. Theor. Phys.} {\bf 37}  649\\
$[32]$   Yan F L and  Bai Y K  2003 {\it Commun. Theor. Phys.} {\bf 40}  273\\
$[33]$  Yan F L and   Wang D 2003 {\it Phys. Lett.} A {\bf  316}  297\\
$[34]$  Gao T, Wang Z X and  Yan F L 2003 {\it Chin. Phys. Lett.} {\bf 20}  2094\\
$[35]$   Gao T,  Yan  F L and  Wang Z X  arXiv: quant-ph/0311141 \\
$[36]$   Gao T,  Yan  F L and  Wang Z X  arXiv: quant-ph/0312021, to published in {\it Commun. Theor. Phys.}\\
$[37]$   Zhou J D, Hou G, Wu S J and Zhang Y D arXiv: quant-ph/0006030\\
$[38]$   Bouwmeester D  et al 1997  {\it Nature} {\bf 390}  575\\
$[39]$   Boschi D  et al 1998 {\it Phys. Rev. Lett.} {\bf 80}  1121\\
$[40]$    Nielsen M A  et al 1998 {\it Nature}  {\bf 396}  52\\
\end{tabbing}

 \end{document}